\documentclass[aps,pre,twocolumn,showpacs]{revtex4}
\usepackage{epsfig}
\usepackage{times}
\usepackage{amsmath}
\begin{document}

\title{Aspiring to the fittest and promotion of cooperation in the prisoner's dilemma game}

\author{Zhen Wang$^1$ and Matja{\v z} Perc$^{2,}$\footnote{Corresponding author.\\Electronic address: matjaz.perc@uni-mb.si}}
\affiliation
{$^1$School of Physics, Nankai University, Tianjin 300071, China\\
$^2$Department of Physics, Faculty of Natural Sciences and Mathematics, University of\\ Maribor, Koro{\v s}ka cesta 160, SI-2000 Maribor, Slovenia}

\begin{abstract}
Strategy changes are an essential part of evolutionary games. Here we introduce a simple rule that, depending on the value of a single parameter $w$, influences the selection of players that are considered as potential sources of the new strategy. For positive $w$ players with high payoffs will be considered more likely, while for negative $w$ the opposite holds. Setting $w$ equal to zero returns the frequently adopted random selection of the opponent. We find that increasing the probability of adopting the strategy from the fittest player within reach, \textit{i.e.} setting $w$ positive, promotes the evolution of cooperation. The robustness of this observation is tested against different levels of uncertainty in the strategy adoption process and for different interaction network. Since the evolution to widespread defection is tightly associated with cooperators having a lower fitness than defectors, the fact that positive values of $w$ facilitate cooperation is quite surprising. We show that the results can be explained by means of a negative feedback effect that increases the vulnerability of defectors although initially increasing their survivability. Moreover, we demonstrate that the introduction of $w$ effectively alters the interaction network and thus also the impact of uncertainty by strategy adoptions on the evolution of cooperation.
\end{abstract}

\pacs{02.50.Le, 87.23.-n, 89.65.-s}
\maketitle

\section{Introduction}

Cooperation within groups of selfish individuals is ubiquitous in human and animal societies. To explain and understand the origin of this phenomenon, evolutionary games, providing a suitable theoretical framework, have been studied extensively by many researches from various disciplines over the past decades \cite{maynard_82, weibull_95, nowak_06}. The evolutionary prisoner's dilemma game in particular, illustrating the social conflict between cooperative and selfish behavior, has attracted considerable attention both in theoretical as well as experimental studies \cite{axelrod_84}. In a typical prisoner's dilemma \cite{hofbauer_98}, two players simultaneously decide whether they wish to cooperate or defect. They will receive the reward R if both cooperate, and the punishment P if both defect. However, if one player defects while the other decides to cooperate, the former will get the temptation T while the latter will get the sucker's payoff S. The ranking of these four payoffs is T$>$R$>$P$>$S, from where it is clear that players need to defect if they wish to maximize their own payoff, irrespective of the opponent's decision. Resulting is a social dilemma, which typically leads to widespread defection. To overcome this unfortunate outcome, several mechanisms that support the evolution of cooperation have been identified (see \cite{nowak_s06} for a review).

Of particular renown are the investigations of spatial prisoner's dilemma games, which have turned out to be very inspirational over decades. In the first spatial prisoner's dilemma game introduced by Nowak and May \cite{nowak_n92b}, players were located on a square lattice, and their payoffs were gathered from the games with their neighbors. Subsequently, players were allowed to adopt the strategy of their neighbors, providing their fitness was higher. It was shown that the introduction of spatial structure enables cooperators to form clusters, thereby promoting the evolution of cooperation. Along this pioneering line of research, many different mechanisms aimed at sustaining cooperation were subsequently proposed and investigated. Examples include the reward mechanism \cite{jimenez_jtb08, cuesta_jtb08}, simultaneous adoption of different strategies depending on the opponents \cite{wardil_epl09}, preferential selection of a neighbor \cite{wu_zx_pre06, fu_pre08, van-segbroeck_prl09, chen_xj_pre09b, shi_dm_pa09}, the mobility of players \cite{vainstein_pre01, vainstein_jtb07, helbing_acs08, helbing_pnas09, meloni_pre09, droz_epjb09, wu_zx_pre09, sicardi_jtb09, jiang_ll_pre10}, heterogeneous teaching activity \cite{szolnoki_epl07, szabo_pre09}, differences in evolutionary time scales \cite{roca_prl06, wu_zx_pre09b}, neutral evolution \cite{cremer_njp09}, and coevolutionary selection of dynamical rules \cite{szolnoki_pre09d, szabo_epl09}, to name but a few. Looking at some examples more specifically, in a recent research paper \cite{fu_pre09}, where players were allowed to either adjust their strategy or switch their defective partners, an optimal state that maximizes cooperation was reported. In \cite{helbing_acs08, helbing_pnas09} it was shown that the mobility of players can lead to an outbreak of cooperation, even if the conditions are noisy and don't necessarily favor the spreading of cooperators. Inspired by these successful research efforts, an interesting question posses itself, which we aim to address in what follows. Namely, if we consider a simple addition to the prisoner's dilemma game that allows players to aspire to the fittest, \textit{i.e.} introducing the propensity of designating the most successful neighbor as being the role model, is this beneficial for the evolution of cooperation or not? The answer is not straightforward since, as we have mentioned, defectors spread by means of their higher fitness. Thus, the modification we consider might give them higher chances of replication. In the early pioneering works, Nowak et al. \cite{nowak_pnas94, nowak_ijbc94} have shown that increasing the probability to copy high payoff neighbors asymptotically leads to increased cooperation, yet this dependence was not monotonic over the whole parameter range. Here we aim to investigate this further in the presence of different levels of uncertainty by strategy adoptions and provide an interpretation of reported results.

Aside from the progress in promoting cooperation described above, another very important development came from replacing the square lattice with more complex interaction topologies (see \cite{szabo_pr07} for a review), possibly reflecting the actual state in social networks more closely. Recently, many studies have attested to the fact that complex networks play a critical role in the maintenance of cooperation for a wide range of parameters \cite{abramson_pre01, santos_prl05, vukov_pre06, santos_prslb06, hauert_ajp05, rong_pre07, gomez-gardenes_prl07, pusch_pre08, poncela_eplp09, perc_njp09}. Quite remarkably, in the early investigations, it has been discovered that the scale-free network can greatly elevate the survivability of cooperators if compared to the classical square lattice \cite{santos_prl05}. Following this discovery, many studies have built on it in order to extend the scope of cooperation on complex networks. For example, a high value of the clustering coefficient was found beneficial \cite{assenza_pre08}, while payoff normalization was found to impair the evolution of cooperation \cite{tomassini_ijmpc07, masuda_prsb07, szolnoki_pa08}. Motivated by these studies, we examine also how aspiring to the fittest in the prisoner's dilemma game fares on complex networks; in particular, whether it promotes or hinders the evolution on cooperation.

Here we thus study the prisoner's dilemma game with the introduction of a mechanism that allows players to aspire to the fittest. Comparing with previous works \cite{szabo_pre98, hauert_ajp05}, where a neighbor was chosen uniformly at random from all the neighbors, the propensity of designating the most successful neighbor as the role model is the most significant difference. Our aim is to study how this mechanism affects the evolution of cooperation on the square lattice, as well as on the scale-free network and the random regular graph, for different levels of uncertainty by strategy adoptions. By means of systematic computer simulations we demonstrate, similarly as was reported already by Nowak et al. \cite{nowak_pnas94, nowak_ijbc94}, that this simple mechanism can actually promote the evolution of cooperation significantly. We give an interpretation of the observed phenomena and examine the impact of different levels of uncertainty by strategy adoptions and the impact of different interaction networks on the outcome of the modified prisoner's dilemma. In the remainder of this paper we will first describe the considered evolutionary game, subsequently we will present the main results, and finally we will summarize our conclusions.

\section{Evolutionary game}

We consider an evolutionary prisoner's dilemma game with the temptation to defect $T = b$ (the highest payoff received by a defector if playing against a cooperator), reward for mutual cooperation $R = b-c$, the punishment for mutual defection $P=0$, and the sucker's payoff $S=-c$ (the lowest payoff received by a cooperator if playing against a defector). For positive $b>c$ we have $T>R>P>S$, thus strictly satisfying the prisoner's dilemma payoff ranking. For simplicity, but without loss of generality, the payoffs can be rescaled such that $R=1$, $T=1+r$, $S=-r$ and $P=0$, where $r=c/(b-c)$ is the cost-to-benefit ratio \cite{hauert_ajp05}. Depending on the interaction network, the strategy adoption rule and other simulation details (see \textit{e.g.} \cite{szabo_pr07, roca_plr09, perc_bs10}), there always exists a critical cost-to-benefit ratio $r=r_c$ at which cooperators die out. We will be interested in determining to what extend does aspiring to the fittest, as we are going to introduce in what follows, affects this critical value under different circumstances.

Throughout this work each player $x$ is initially designated either as a cooperator ($s_x=$C) or defector (D) with equal probability. As the interaction network, we use either a regular $L \times L$ square lattice, the random regular graph constructed as described in \cite{szabo_jpa04}, or the scale-free network with $L^2$ nodes and an average degree of four generated via the Barab{\'a}si-Albert algorithm \cite{barabasi_s99}. The game is iterated forward in accordance with the sequential simulation procedure comprising the following elementary steps. First, player $x$ acquires its payoff $p_x$ by playing the game with all its neighbors. Next, we evaluate in the same way the payoffs of all the neighbors of player $x$ and subsequently select one neighbor $y$ via the probability
\begin{equation}
\Pi_{y}=\frac{\exp(w p_{y})}{\sum_{z} \exp(w p_{z})},
\end{equation}
where the sum runs over all the neighbors of player $x$ and $w$ is the newly introduced selection parameter. Evidently, for $w=0$ the most frequently adopted situation is recovered where player $y$ is chosen uniformly at random from all the neighbors of player $x$. For $w>0$, however, Eq.~(1) introduces a preference towards those neighbors of player $x$ that have a higher payoff $p_y$. Conversely, for $w<0$ players with a lower payoff are more likely to be selected as potential strategy donors. Lastly then, player $x$ adopts the strategy $s_y$ from the selected player $y$ with the probability
\begin{equation}
W(s_y \rightarrow s_x)=\frac{1}{1+\exp[(p_x-p_y)/K]},
\end{equation}
where $K$ denotes the amplitude of noise or its inverse ($1/K$) the so-called intensity of selection \cite{szabo_pre98}. Irrespective of the value of $w$ one full iteration step involves all players $x=1,2, \ldots, L^2$ having a chance to adopt a strategy from one of their neighbors once. Here the evolutionary prisoner's dilemma game is thus supplemented by a selection parameters $w$, enabling us to tune the preference towards which neighbor will be considered more likely as a potential strategy donor. For positive values of $w$ the players are more likely to aspire to their most fittest neighbors, while for negative values of $w$ the less successful neighbors will more likely act as strategy donors. This amendment seems reasonable and is easily justifiable with realistic examples. For example, it is a fact that people are, in general, much more likely to follow a successful individual than someone who is struggling to get by. This is taken into account by positive values of $w$. However, under certain (admittedly rare) circumstances, it is also possible that individuals will be inspired to copy their less successful partners. Indeed, the most frequently adopted random selection of a neighbor, retrieved in our case by $w=0$, seems in many ways like the least probable alternative. It is also informative to note that aspiring to the fittest becomes identical to the frequently adopted ``best takes all'' rule if $w \to \infty$ in Eq.~(1) and $K \to 0$ in Eq.~(2). This rule was adopted in the seminal work by Nowak and May \cite{nowak_n92b}, as well as subsequently by Huberman and Glance \cite{huberman_pnas93} who showed that under certain circumstances asynchronous updating is substantially less successful in ensuring the survivability of cooperators than synchronous updating. Although in our simulations we never quite reach the ``best takes all'' limit, and thus a direct comparison is somewhat circumstantial, it is interesting to note that an additional uncertainty in the strategy adoption process via finite values of $K$ may alleviate the disadvantage that is due to asynchronous updating \cite{szabo_pre98}.

Results of computer simulations presented below were obtained on populations comprising $100 \times 100$ to $400 \times 400$ individuals, whereby the fraction of cooperators $\rho_{{\rm C}}$ was determined within $10^5$ full iteration steps after sufficiently long transients were discarded. Moreover, since the preferential selection of neighbors may introduce additional disturbances, final results were averaged over up to $40$ independent runs for each set of parameter values in order to assure suitable accuracy.

\section{Results}

\begin{figure}
\centerline{\epsfig{file=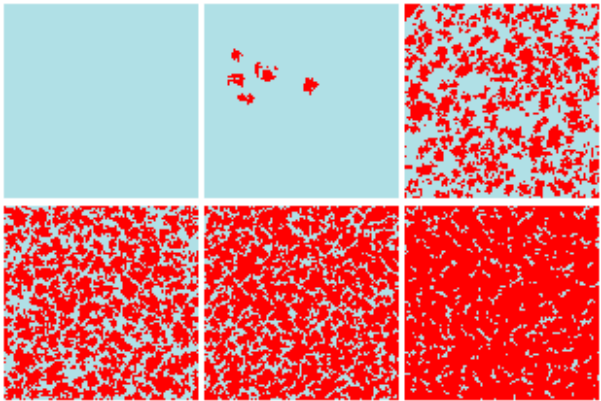,width=8cm}}
\caption{(\textit{color online}) Characteristic snapshots of cooperators [red (dark gray in black-white print)] and defectors [light blue (light gray in black-white print)] for different values of the selection parameter $w$. From top left to bottom right $w = -0.2$, $0$, $0.2$, $0.5$, $1.0$ and $4.0$, respectively. All panels depict results obtained for $r=0.022$ and $K=0.1$ on a $100 \times 100$ square lattice.}
\label{fig1}
\end{figure}

We start by visually inspecting characteristic spatial distributions of cooperators and defectors for different values of the selection parameter $w$. Figure~\ref{fig1} features the results obtained for $r=0.022$ and $K=0.1$, whereat for $w=0$ (upper middle panel) a small fraction of cooperators can prevail on the square lattice by means of forming clusters, thereby protecting themselves against the exploitation by defectors \cite{hauert_n04}. As evidenced in the upper leftmost panel, for negative values of $w$ even this small fraction of cooperators goes extinct, thus yielding as a results exclusive dominance of defectors. For positive values of $w$ (upper right panel), however, the cooperators start mushrooming, whereby clustering remains their mechanism of spreading and survivability. Interestingly, large enough values of $w$ can facilitate the evolution of cooperation to the point of near-complete cooperator dominance (bottom right panel), or at least equality with the defectors, as implied by $\rho_{{\rm C}} \geq \rho_{{\rm D}}$ in all lower panels of Fig.~\ref{fig1}. These results suggest that when players aspire to adopt the strategy from their fittest neighbor the evolution of cooperation thrives. In what follows we will systematically examine the validity of this claim.

\begin{figure}
\centerline{\epsfig{file=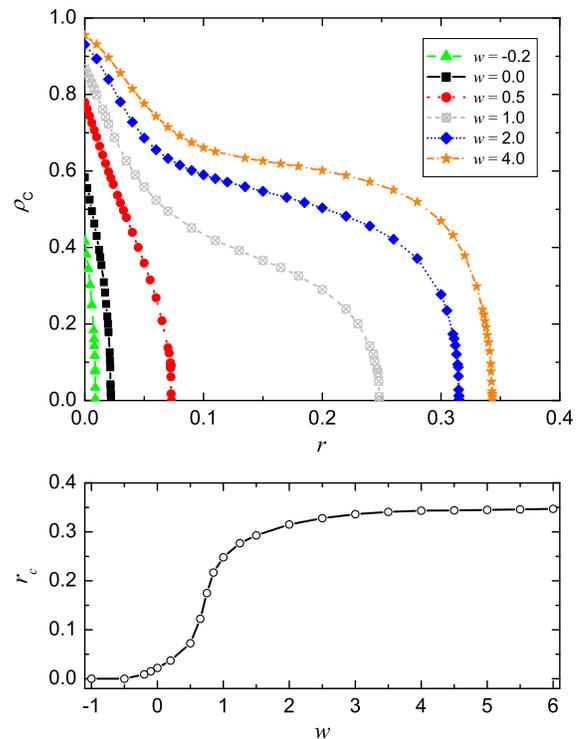,width=7.6cm}}
\caption{(\textit{color online}) \textit{Top panel}: Frequency of cooperators $\rho_{{\rm C}}$ in dependence on the cost-to-benefit ratio $r$ for different values of the selection parameter $w$. From left to right $w=-0.2$, $0$, $0.5$, $1.0$, $2.0$ and $4.0$, respectively. Note that negative values of $w$ impair the evolution of cooperation, while $w>0$ move the survivability of cooperators towards larger values or $r$. \textit{Bottom panel}: Critical threshold values of the cost-to-benefit ratio $r=r_{c}$, marking the transition to the pure D phase (extinction of cooperators), in dependence on the selection parameter $w$. Note that $r_{c}$ converges in both the negative and the positive limit of $w$. In particular, $r_{c} \to 0$ for negative and $r_{c} \to 0.35$ for positive values of $w$. Depicted results were obtained for $K=0.1$ (both panels).}
\label{fig2}
\end{figure}

To quantify the ability of particular values of the selection parameter to facilitate and maintain cooperation more precisely, we first calculate $\rho_{{\rm C}}$ in dependence on the cost-to-benefit ratio $r$ for different values of $w$. Results presented in the top panel of Fig.~\ref{fig2} clearly attest to the fact that positive values of $w$ promote the evolution of cooperation, while on the other hand, negative values of $w$ impede it. Note that the critical cost-to-benefit ratio $r=r_{c}$, marking the extinction of cooperators, increases by a full order of magnitude at $w=4.0$ (orange stars) if compared to the $w=0$ (black squares) case. Interestingly, the promotive effect on the survivability of cooperators becomes more potent monotonously with increasing $w$, thus suggesting that a universally applicable mechanism is underlying the observed behavior. Indeed, the monotonous increase of $r=r_{c}$ for increasing $w$ is obvious from the bottom panel of Fig.~\ref{fig2}, showing concisely the extend to which aspiring to the fittest promotes the evolution of cooperation on the square lattice.

\begin{figure}
\centerline{\epsfig{file=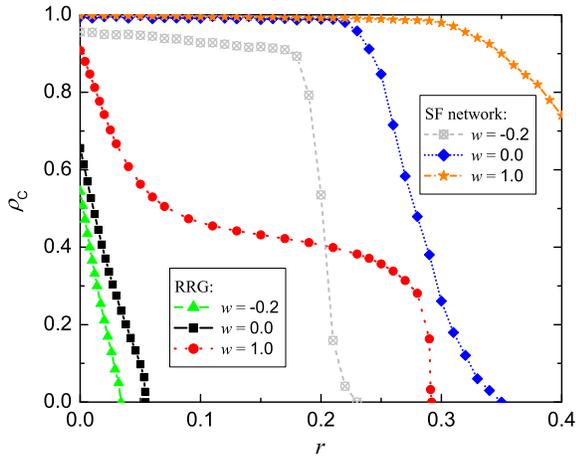,width=7.7cm}}
\caption{(\textit{color online}) Frequency of cooperators $\rho_{{\rm C}}$ in dependence on the cost-to-benefit ratio $r$ for different values of the selection parameter $w$ for the random regular graph (RRG) and the scale-free (SF) network. From left to right $w=-0.2$, $0$, $1.0$ for the random regular graph, and $w=-0.2$, $0$, $1.0$ for the scale-free network, respectively. Note that these results are in qualitative agreement with those obtained on the square lattice in that negative values of $w$ impair the evolution of cooperation, while $w>0$ move the survivability of cooperators towards larger values or $r$. Depicted results were obtained for $K=0.1$.}
\label{fig3}
\end{figure}

Importantly, qualitatively identical results can be obtained on interaction networks other than the square lattice. Results presented in Fig.~\ref{fig3} depict how cooperators fare on the random regular graph and the scale-free network for different values of $w$. Similarly as in Fig.~\ref{fig2}, it can be observed that positive values of $w$ promote the evolution of cooperation. Conversely, negative values of $w$ promote the evolution of defection. This is in agreement with the observations made on the square lattice, thus designating $w>0$ as being universally effective in promoting the evolution of cooperation, in particular, working on regular lattices and graphs as well as highly heterogeneous networks. Since the latter have been identified as potent promoters of cooperation on their own right \cite{santos_prl05}, this conclusion is all the more inspiring.

\begin{figure}
\centerline{\epsfig{file=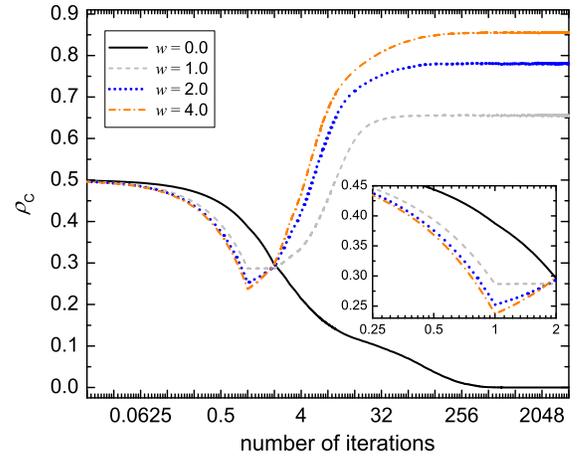,width=7.57cm}}
\caption{(\textit{color online}) Time courses depicting the evolution of cooperation for $w=0$ (solid black line), $w=1.0$ (dashed gray line), $w=2.0$ (dotted blue line) and $w=4.0$ (dash-dotted orange line). Note that while for $w=0$ cooperators die out, for $w>0$ they recover from what appears to become an even faster extinction to eventually rise to near-dominance. Notably, the stronger the initial temporary downfall, the better the recovery (see also the inset). All time courses were obtained as averages over $20$ independent realizations for $r=0.03$ and $K=0.1$ on a $400 \times 400$ square lattice. Note that the horizontal axis is logarithmic and that values of $\rho_{{\rm C}}$ were recorded also in-between full iteration steps to ensure a proper resolution.}
\label{fig4}
\end{figure}

In order to explain the promotive impact of positive values of $w$ on the evolution of cooperation, we examine time courses of $\rho_{{\rm C}}$ for different values of the selection parameter. Figure~\ref{fig4} features results obtained for $r=0.03$ and $K=0.1$, whereat cooperators die out if $w=0$ (black line; see also Fig.~\ref{fig2}). For positive values of $w$, on the other hand, the stationary state is a mixed C+D phase with cooperators occupying the larger portion of the square lattice. Interestingly, however, in the most early stages of the evolutionary process (note that values of $\rho_{{\rm C}}$ were recorded also in-between full iteration steps) it appears as if defectors would actually fare better for $w>0$. In fact, the larger the value of $w$, the deeper the initial downfall of cooperators. This is actually what one would expect, given that defectors are, as individuals, more successful than cooperators and will thus be chosen more likely as potential strategy donors if $w$ is positive. This in turn amplifies their chances of spreading and results in the decimation of cooperators (only slightly more than 20~\% survive). Quite surprisingly though, the tide changes fast, and as one can observe from the presented time courses, the more so the deeper the initial downfall of cooperators. For $w=4.0$ we can observe instead of cooperator extinction their near-dominance with $\rho_{{\rm C}}$ hoovering comfortably over $0.8$ (orange line). We argue that for positive values of $w$ a negative feedback effect occurs, which halts and eventually reverts what appears to be a march of defectors towards their undisputed dominance. Namely, in the very early stages of the game defectors are able to plunder very efficiently, which quickly results in a state where there are hardly any cooperators left to exploit. Consequently, the few remaining clusters of cooperators start recovering lost ground against weakened defectors. Crucial thereby is the fact that the clusters formed by cooperators are impervious to defector attacks even at high values of $r$ because of the positive selection towards the fittest neighbors acting as strategy sources (occurring for $w>0$). In a sea of cooperators this is practically always another cooperator rather than a defector trying to penetrate into the cluster. This newly identified mechanism ultimately results in widespread cooperation that goes beyond what can be warranted by the spatial reciprocity alone (see \textit{e.g.} \cite{szabo_pr07}), and this irrespective of the underlying interaction network. As such, aspiration to the fittest, \textit{i.e.} the propensity of designating the most successful neighbor as being the role model, may be seen as a universally applicable promoter of cooperation.

\begin{figure}
\centerline{\epsfig{file=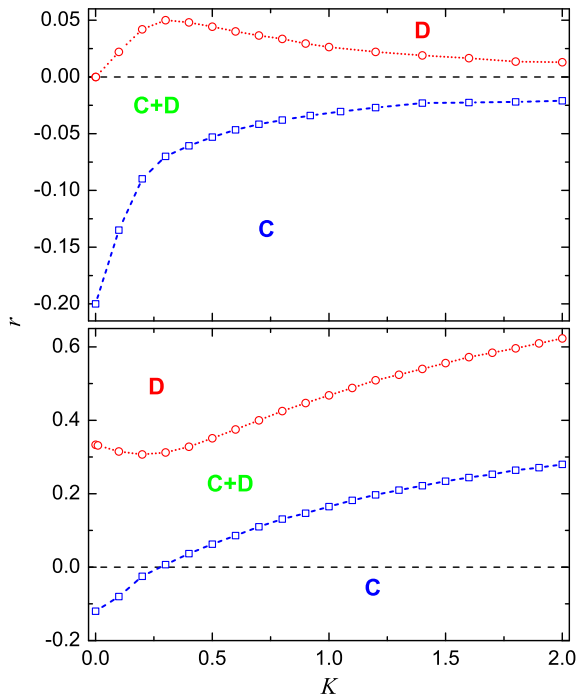,width=7.7cm}}
\caption{Full $r-K$ phase diagram for $w=0$ (top panel) and $w=2.0$ (bottom panel), obtained via systematic simulations of the prisoner's dilemma game on the square lattice. Dashed blue and dotted red lines mark the border between stationary pure C and D phases and the mixed C+D phase, respectively. In agreement with previous works \cite{szabo_pre05, vukov_pre06}, it can be observed that for $w=0$ (top panel) there exists an intermediate uncertainty in the strategy adoption process (an intermediate value of $K$) for which the survivability of cooperators is optimal, \textit{i.e.} $r_{c}$ is maximal. Conversely, while the borderline separating the pure C and the mixed C+D phase for the $w=2.0$ case (bottom panel) exhibits a qualitatively identical outlay as for the $w=0$ case, the D $\leftrightarrow$ C+D transition is qualitatively different. Note that in the bottom panel there exist an intermediate value of $K$ for which $r_{c}$ is minimal rather than maximal, while towards the large $K$ limit $r_{c}$ increases, saturating only for $K>4$ (not shown).}
\label{fig5}
\end{figure}

Lastly, it is instructive to examine the evolution of cooperation for $w>0$ in dependence on the uncertainty by strategy adoptions. The latter can be tuned via $K$, which acts as a temperature parameter in the employed Fermi strategy adoption function \cite{szabo_pre98}. Accordingly, when $K \to \infty$ all information is lost and the strategies are adopted by means of a coin toss. The phase diagram presented in the top panel of Fig.~\ref{fig5} is well-known, implying the existence of an optimal level of uncertainty for the evolution of cooperation, as was previously reported in \cite{perc_njp06a, vukov_pre06}. In particular, note that the D $\leftrightarrow$ C+D transition line is bell shaped, indicating that $K \approx 0.37$ is the optimal temperature at which cooperators are able to survive at the highest value of $r$. This phenomenon can be interpreted as an evolutionary resonance \cite{perc_njp06b}, albeit it can only be observed on interaction topologies lacking overlapping triangles \cite{szabo_pre05, szolnoki_pre09c}. Interestingly, positive $w$ eradicate (as do interaction networks incorporating overlapping triangles) the existence of an optimal $K$, as can be observed from the phase diagram presented in the bottom panel of Fig.~\ref{fig5}. The latter was obtained for $w=2.0$ and exhibits an inverted bell-shaped D $\leftrightarrow$ C+D transition line, indicating the existence of the worst rather than an optimal temperature $K$ for the evolution of cooperation. This in turn implies that introducing a preference towards the fittest neighbors effectively alters the interaction network. While the square lattice obviously lacks overlapping triangles and thus enables the observation of an optimal $K$, trimming the likelihood of who will act as a strategy source seems to effectively enhance linkage among essentially disconnected triplets and thus precludes the same observation. A similar phenomenon was observed recently in public goods games, where the joint membership in large groups was also found to alter the effective interaction network and thus the impact of uncertainly on the evolution of cooperation \cite{szolnoki_pre09c}.

\section{Summary}

In sum, we have shown that aspiring to the fittest promotes the evolution of cooperation in the prisoner's dilemma game irrespective of the underlying interaction network and the uncertainty by strategy adoptions. The essence of the identified mechanism for the cooperation promotion has been attributed to a negative feedback effect, occurring because of the formation of extremely robust clusters (or groups on complex networks) of cooperators that are impervious to defector attacks even at high temptations to defect. Although initially the defectors appear to be heading to an undisputed victory, the fast exploitation and the consequent shortage of cooperators weakens the defectors and makes them susceptible to an overtake by the few remaining cooperators. Further interesting is the fact that the introduction of a selection parameter, making the fittest neighbors more likely to act as sources of adopted strategies, effectively alters the interaction network. While in its absence there exists an intermediate uncertainty governing the process of strategy adoptions $K$ by which the largest cost-to-benefit ratio $r$ still warrants the survival of at least some cooperators, in its presence this feature vanishes and becomes qualitatively identical to what was observed previously on lattices that do incorporate overlapping triangles, such as the kagome lattice \cite{szolnoki_pre09c}. Since in fact the actual interaction topology remains unaffected by the different values of the selection parameter $w$, we have argued that the differences in the evolution of cooperation are due to an effective transition of the interaction topology, which is brought about by the fact that some players are more likely to act as strategy sources than others. Therefore, the bonds between certain player pairs appears stronger than average, although the interaction networks consist of links that are not weighted.

Since aspiring to the fittest, \textit{i.e.} the propensity of designating the most successful neighbors as role models, appears to be both widely applicable as well realistically justifiable, we hope it will inspire future studies, especially in terms of understanding the emergence of successful leaders in societies via a coevolutionary process \cite{perc_bs10}. An interesting interpretation of the selection parameter $w$ can also be obtained if the latter is considered as a measure of cognitive complexity of each individual. In particular, it is possible to argue that the more obtuse an individual is, the closer to random his choice of a role model will be. If individuals are to be able to aspire to the fittest, they should have some degree of information processing capabilities. On the other hand, negative values of $w$ can be interpreted as a choice that is based on moral values \cite{helbing_plos10}, for example, when highly successful individuals are so by unethical actions and thus should not be imitated.

\begin{acknowledgments}
Matja{\v z} Perc acknowledges support from the Slovenian Research Agency (Grant No. Z1-2032). Zhen Wang acknowledges support from the Center for Asia Studies of Nankai University (Grant No. 2010-5) and from the National Natural Science Foundation of China (Grant No. 10672081). Helpful discussion with Professor Lianzhong Zhang are gratefully acknowledged as well. This works has benefited substantially from the insightful comments of the Physical Review referees, and we are very grateful for their help.
\end{acknowledgments}

\end{document}